\documentclass[twocolumn,english,showpacs,pre]{revtex4}
\usepackage{amsmath}
\usepackage{graphicx}
\usepackage{amssymb}
\usepackage{babel}

\begin{document}

\title{Spatial distribution of the emission intensity in a photonic crystal:
Self-interference of Bloch eigenwaves}

\author{Dmitry N. Chigrin}

\email{chigrin@th.physik.uni-bonn.de}

\affiliation{Physikalisches Institut, Universit\"at Bonn, Nussallee 12, D-53115
Bonn, Germany}

\begin{abstract}
The analysis of an angular distribution of the emission intensity
of a two-level atom (dipole) in a photonic crystal reveals an enhancement
of the emission rate in some observation directions. Such an enhancement
is the result of the bunching of many Bloch eigenwaves with different
wave vectors in the same direction due to the crystal anisotropy.
If a spatial distribution of the emission intensity is considered,
the interference of these eigenwaves should be taken into account.
In this paper, the far-field emission pattern of a two-level atom
is discussed in the framework of the asymptotic analysis the classical
macroscopic Green function. Numerical example is given for a two-dimensional
square lattice of air holes in polymer. The relevance of results for
experimental observation is discussed.
\end{abstract}


\maketitle

\section{Introduction}

Light emission in a photonic crystal has attracted a substantial attention
both in theoretical \citep{Byk72,eli87,john90,dowling92,suzuki95,sakoda96,nojima98,busch00,xu00,li00,lousse01,hermann02}
and experimental \citep{dowling94,hirayama96,boroditsky99,gaponenko99,temelkuran00,schriemer00,romanov01,koenderink02,romanov03,koenderink03}
studies. To a large extend this interest is due to potential perspectives
of the emission modification and control provided by photonic crystals.
The inhibition of spontaneous emission is possible within a spectral
range of a complete photonic bandgap \citep{Byk72,eli87}, where linear
propagation of light is prohibited in all spatial directions. An emission
enhancement is a result of a long interaction time of an emitter and
the radiated field, when the emitter is coupled to the slow eigenmode
\citep{dowling94,nojima98,sakoda99} or to the strongly localized
mode of a defect state of a photonic crystal \citep{bookJDJ,eli98,painter99}.

It is well known that, the spontaneous decay of an excited atom strongly
depends on the environment \citep{purcell46}. Both the emission rate
and the emission directionality can be affected. In the simplest case
of a two-level atom placed in an inhomogeneous medium the emission
dynamics can be described by the integro-differential equation for
the upper state occupation probability amplitude \citep{busch00,Dung2000}
\begin{equation}
\dot{C}\left(t\right)=-\int_{0}^{t}dt'\, K\left(t-t'\right)C\left(t\right)\label{eq:dynamics}\end{equation}
with the kernel $K\left(t-t'\right)$defined by\begin{equation}
K\left(t-t'\right)=\frac{1}{2\hbar\varepsilon_{0}}\int_{0}^{\infty}d\omega\,\omega\rho\left(\mathbf{r}_{0},\omega\right)e^{-i\left(\omega-\omega_{0}\right)\left(t-t'\right)},\label{eq:kernel}\end{equation}
where\begin{equation}
\rho\left(\mathbf{r}_{0},\omega\right)=\frac{2\omega}{\pi c^{2}}\mathbf{d}\cdot\mathrm{Im}\left[\overleftrightarrow{\mathbf{G}}\left(\mathbf{r}_{0},\mathbf{r}_{0},\omega\right)\right]\cdot\mathbf{d}\label{eq:pldos}\end{equation}
is the projected local density of states (PLDOS) \citep{Sprik1996,McPhedran2004}.
$\overleftrightarrow{\mathbf{G}}\left(\mathbf{r},\mathbf{r}',\omega\right)$
is the classical macroscopic dyadic Green function defined by the
inhomogeneous wave equation \citep{bookSakoda}\begin{equation}
\left(\frac{\omega^{2}}{c^{2}}\varepsilon(\mathbf{r})-\nabla\times\nabla\times\right)\overleftrightarrow{\mathbf{G}}\left(\mathbf{r},\mathbf{r}',\omega\right)=-\overleftrightarrow{\delta}_{\epsilon_{\perp}}\left(\mathbf{r}-\mathbf{r}'\right).\label{eq:weq}\end{equation}
Here $\overleftrightarrow{\delta}_{\epsilon_{\perp}}\left(\mathbf{r}-\mathbf{r}'\right)$
is the $\varepsilon$-transverse dyadic delta function \citep{glauber91,dowling92},
$\mathbf{r}_{0}$, $\omega_{0}$ and $\mathbf{d}$ are atom location,
transition frequency and transition dipole moment, respectively. $c$
is a speed of light in vacuum. In the case of a general linear, non-magnetic,
dielectric medium with arbitrary three-dimensional (3D) periodic dielectric
function, $\varepsilon(\mathbf{r})$, the Green's function can be
expressed in the Bloch mode basis as \citep{dowling92,bookSakoda}\begin{equation}
\overleftrightarrow{\mathbf{G}}\left(\mathbf{r},\mathbf{r}',\omega\right)=\frac{c^{2}}{V}\sum_{n\mathbf{k}}\frac{\mathbf{A}_{n\mathbf{k}}\left(\mathbf{r}\right)\otimes\mathbf{A}_{n\mathbf{k}}^{*}\left(\mathbf{r}'\right)}{\omega_{n\mathbf{k}}^{2}-\left(\omega+i\gamma\right)^{2}}\label{eq:G(r,r,w)}\end{equation}
Here $\mathbf{A}_{n\mathbf{k}}(\mathbf{r})$ are Bloch eigenwaves
characterized by the band index $n$, the wave vector $\mathbf{k}_{n}$
and the eigenfrequencies $\omega_{n\mathbf{k}}$. Bloch eigenwaves
are solutions of the homogeneous wave equation and obey the gauge
$\nabla\cdot\left[\varepsilon(\mathbf{r})\mathbf{A}_{n\mathbf{k}}(\mathbf{r})\right]=0$,
normalization and completeness conditions \citep{glauber91,dowling92}.
The asterisk $*$ and $\otimes$ denote the complex conjugate and
the outer tensor product, respectively. A positive infinitesimal $\gamma$
assures causality.

In the limit of the weak coupling (Markov approximation), a coarse-grained
description of the atomic motion memory effects can be disregarded
\citep{busch00,Dung2000} and Eq. (\ref{eq:dynamics}) yields the
familiar exponential decay of the excited state with a decay rate
given by\begin{eqnarray}
\Gamma\left(\omega_{0}\right) & = & \frac{\pi\omega_{0}}{\hbar\varepsilon_{0}}\rho\left(\mathbf{r}_{0},\omega_{0}\right)\nonumber \\
 & = & \frac{\pi\omega_{0}}{\hbar\varepsilon_{0}V}\sum_{n\mathbf{k}}\left|\mathbf{A}_{n\mathbf{k}}\left(\mathbf{r}_{0}\right)\cdot\mathbf{d}\right|^{2}\delta\left(\omega_{0}-\omega_{n\mathbf{k}}\right)\label{eq:gamma}\end{eqnarray}
and the spatial distribution of the emitted intensity given by \citep{Dung2000}
\begin{eqnarray}
I\left(\mathbf{r,\omega}_{0}\right) & = & \left|\frac{\omega_{0}^{2}}{\varepsilon_{0}c^{2}}\overleftrightarrow{\mathbf{G}}\left(\mathbf{r},\mathbf{r}_{0},\omega_{0}\right)\cdot\mathbf{d}\right|^{2}\nonumber \\
 & = & \left|\frac{\omega_{0}^{2}}{\varepsilon_{0}V}\sum_{n\mathbf{k}}\frac{\left(\mathbf{A}_{n\mathbf{k}}^{*}\left(\mathbf{r}_{0}\right)\cdot\mathbf{d}\right)\mathbf{A}_{n\mathbf{k}}\left(\mathbf{r}\right)}{\omega_{n\mathbf{k}}^{2}-\left(\omega_{0}+i\gamma\right)^{2}}\right|^{2}.\label{eq:I(r,w)}\end{eqnarray}
This approximation gives the correct result for emission modification
in most of the situations considered in the present paper. In the
same time, special care should be taken for frequencies near the photonic
band edges or other van Hove singularities, where the memory effects
become significant and Eq.~(\ref{eq:dynamics}) should be analyzed
instead of a direct use of Eqs.~(\ref{eq:gamma},\ref{eq:I(r,w)}).

In Eqs.~(\ref{eq:dynamics}) and (\ref{eq:gamma},\ref{eq:I(r,w)})
all parameters of a periodic environment relevant for the atomic evolution
are contained via the classical Green function and its Bloch mode
expansion (\ref{eq:G(r,r,w)}). The emission rate (\ref{eq:gamma})
is proportional to the density of available Bloch eigenmodes weighed
by the coupling strength between the atomic dipole moment and the
corresponding mode. The emitted intensity (\ref{eq:I(r,w)}) at the
point $\mathbf{r}$ is determined both by the spectral and angular
emission rate modification and by the interference of the Bloch modes
at this point.

In contrast to the total emission rate modification (\ref{eq:gamma}),
angular-resolved emission experiments, e.g. \citep{romanov03,koenderink03},
usually probe only a small fraction of the solid angle detecting the
emission modification in a particular direction in space. To analyzed
such angular-resolved emission experiments, a fractional power emitted
per solid angle in space can introduced. It has been shown, that the
radiation pattern of the classical dipole in a photonic crystal can
demonstrate a strong modification with respect to the dipole radiation
pattern in vacuum \citep{romanov03,chigrin2004}. For example, in
a photonic crystal with incomplete bandgap, the angular-resolved emission
rate is suppressed in the direction of the spatial stopband and strongly
enhanced in the direction of the group velocity, which is stationary
with respect to a small variation of the wave vector (photon focusing)
\citep{chigrin2004}. Such an enhancement is the result of the bunching
of many Bloch eigenwaves with different wave vectors in the same spatial
direction due to the crystal anisotropy \citep{zengerle87,russell86a}.
For a coherent light source, this inevitably leads to the interference
of the Bloch eigenwaves at the detector plane and additional modification
of the spatial distribution of the emission intensity.

In this paper, the physical picture of the interference fringes formation
in the far-field emission pattern of the two-level atom placed in
a periodic medium is considered. The paper is organized as follows.
The evaluation of the asymptotic form of the emitted intensity (\ref{eq:I(r,w)})
is given in section \ref{sec:self:asymptotic} in the radiation zone.
The physical explanation and the relevance of results for experimental
observation are discussed in section \ref{sec:self:physics}. In section
\ref{sec:self:polymer} a numerical example is given for a two-dimensional
polymer photonic crystal. Section \ref{sec:self:summary} summarized
the main results of the paper.

\section{Asymptotic form of emitted intensity\label{sec:self:asymptotic}}

By taking into account the Bloch theorem, $\mathbf{A}_{n\mathbf{k}}(\mathbf{r})=\mathbf{a}_{n\mathbf{k}}(\mathbf{r})e^{i\mathbf{k}_{n}\cdot\mathbf{r}}$,
where $\mathbf{a}_{n\mathbf{k}}(\mathbf{r})$ is a lattice periodic
function, and changing the $k$-space summation to the corresponding
$k$-space integral, $\sum_{\mathbf{k}}\rightarrow\left(V/8\pi^{3}\right)\int d^{3}k$,
Eq. (\ref{eq:I(r,w)}) can be expressed as\begin{widetext} \begin{equation}
I(\mathbf{r},\omega)=\left|\frac{\omega_{0}^{2}}{8\pi^{3}\varepsilon_{0}}\sum_{n}\int d^{3}\mathbf{k}_{n}\frac{\left(\mathbf{a}_{n\mathbf{k}}^{*}\left(\mathbf{r}_{0}\right)\cdot\mathbf{d}\right)\mathbf{a}_{n\mathbf{k}}\left(\mathbf{r}\right)}{\omega_{n\mathbf{k}}^{2}-\left(\omega_{0}+i\gamma\right)^{2}}e^{i\mathbf{k}_{n}\cdot(\mathbf{r}-\mathbf{r}_{0})}\right|^{2}\label{eq:I2}\end{equation}
For large $\mathbf{\left|x\right|}=\left|\mathbf{r}-\mathbf{r_{0}}\right|$
the exponential function in the integral (\ref{eq:I2}) oscillates
rapidly. To evaluate the integral, the method of stationary phase
can be used. As it was shown in \citep{chigrin2004}, the principal
contribution to the integral comes from the regions of the iso-frequency
surface in the wave vector space, at which the eigenwave group velocity
is parallel to observation direction $\mathbf{x}$. Then, the integral
in Eq. (\ref{eq:I2}) can be transformed to the form \begin{equation}
I(\mathbf{r},\omega)\approx\left|\frac{\omega}{8\pi^{2}\varepsilon_{0}}\sum_{\nu}\sum_{n}\frac{\left(\mathbf{a}_{n\mathbf{k}}^{\nu*}(\mathbf{r}_{0})\cdot\mathbf{d}\right)\mathbf{a}_{n\mathbf{k}}^{\nu}(\mathbf{r})}{\left|\mathbf{V}_{n\mathbf{k}}^{\nu}\right|}\oint_{-\infty}^{\infty}d^{2}\mathbf{k}_{n}e^{i\mathbf{k}_{n}\cdot(\mathbf{r}-\mathbf{r}_{0})}\right|^{2},\label{math:self:FnkAsymptotic2D}\end{equation}
where $\mathbf{V}_{n\mathbf{k}}^{\nu}$ is the group velocity of the
Bloch eigenwave, the integration is over the iso-frequency surface
$\omega_{n\mathbf{k}}=\omega_{0}$ and summation is taken over all
stationary eigenwaves $\nu$ with a group velocity vector pointing
in the observation direction $\mathbf{x}=\mathbf{r}-\mathbf{r_{0}}$.

The result of the integration in (\ref{math:self:FnkAsymptotic2D})
depends on the local topology of iso-frequency surface $\omega_{n\mathbf{k}}=\omega_{0}$.
It is convenient to introduce the local curvilinear coordinates $\xi_{i}$
with the origin at the iso-frequency surface and with one of the coordinate
aligned perpendicular to it, e.g., $\xi_{3}$. Then, a function $h\left(\xi_{1},\xi_{2}\right)=\mathbf{k}_{n}\cdot\widehat{\mathbf{x}}$
can be expanded in a series near the wave vector of the eigenwave
$(\nu,n,\mathbf{k})$:

\begin{equation}
h\left(\xi_{1},\xi_{2}\right)=\mathbf{k}_{n}^{\nu}\cdot\widehat{\mathbf{x}}+\frac{1}{2}\sum_{i,j=1}^{2}\alpha_{ij}^{\nu}\xi_{i}\xi_{j}+\frac{1}{6}\sum_{i,j,k=1}^{2}\beta_{ijk}^{\nu}\xi_{i}\xi_{j}\xi_{k}+O\left(\xi_{1},\xi_{2}\right)^{4},\label{math:self:surfaceExpansion}\end{equation}
 where\[
\alpha_{ij}^{\nu}=\left(\frac{\partial^{2}h}{\partial\xi_{i}\partial\xi_{j}}\right)_{\nu},\qquad\beta_{ijk}^{\nu}=\left(\frac{\partial^{3}h}{\partial\xi_{i}\partial\xi_{j}\partial\xi_{k}}\right)_{\nu}\]
 and $\widehat{\mathbf{x}}$ is a unit vector in the observation direction
$\mathbf{x}=\mathbf{r}-\mathbf{r_{0}}$. If the iso-frequency surface
has a non-vanishing Gaussian curvature in the vicinity of the wave
vector $\mathbf{k}_{n}^{\nu}$, only quadratic terms in the expansion
(\ref{math:self:surfaceExpansion}) can be kept, leading to the following
asymptotic from of the far-field intensity(\ref{math:self:FnkAsymptotic2D})
\citep{chigrin2004}\begin{equation}
I(\mathbf{r},\omega)\approx\left|\frac{\omega_{0}^{2}}{4\pi\varepsilon_{0}}\sum_{\nu}\sum_{n}\exp\left(-i\frac{\pi}{4}\left(\mathrm{sign}(\alpha_{1}^{\nu})+\mathrm{sign}(\alpha_{2}^{\nu})\right)\right)\frac{\left(\mathbf{A}_{n\mathbf{k}}^{\nu*}(\mathbf{r}_{0})\cdot\mathbf{d}\right)\mathbf{A}_{n\mathbf{k}}^{\nu}(\mathbf{r})}{\left|\mathbf{V}_{n\mathbf{k}}^{\nu}\right|}\frac{1}{\left|K_{n\mathbf{k}}^{\nu}\right|^{1/2}\left|\mathbf{r}-\mathbf{r}_{0}\right|}\right|^{2}\label{math:self:FnAsymGeometrical}\end{equation}
\end{widetext}where $K_{n\mathbf{k}}^{\nu}=\alpha_{11}^{\nu}\alpha_{22}^{\nu}$
determines the Gaussian curvature of the iso-frequency surface at
the point $\mathbf{k}_{n}=\mathbf{k}_{n}^{\nu}$ (stationary point)
and summation is over all stationary points with $\mathbf{x}\cdot\mathbf{V}_{n\mathbf{k}}^{\nu}>0$.

The emission intensity far from an atom is proportional to the inverse
Gaussian curvature of the iso-frequency surface, $\sim\left|K_{n\mathbf{k}}^{\nu}\right|^{-1}$,
and to the inverse square of the distance between the source and the
observation point, $\sim\left|\mathbf{x}\right|^{-2}$. The asymptotic
energy flux shows the necessary amount of decrease with distance ($\sim\left|\mathbf{x}\right|^{-2}$),
providing a finite value of the energy flux in any finite interval
of a solid angle, that is, assuming non-vanishing Gaussian curvature.
A vanishing curvature formally implies an infinite flux along the
corresponding observation direction, leading to the photon focusing
phenomenon \citep{etchegoin96,chigrin2001,chigrin2004}.

Strictly speaking, the asymptotic behavior of the emission intensity
(\ref{math:self:FnAsymGeometrical}) is valid only if quadratic terms
in the expansion (\ref{math:self:surfaceExpansion}) do not vanish,
so that all higher order terms in the expansion can be neglected.
A parabolic point of the iso-frequency surface is an example of vanishing
quadratic terms in (\ref{math:self:surfaceExpansion}). Generally,
the Gaussian curvature is zero at a parabolic point and one (both)
of the principal curvatures of the iso-frequency surface is zero.
Actually, at parabolic points the asymptotic behavior of the emitted
intensity, i.e., the dependence of the intensity on the inverse distance
($\sim\left|\mathbf{x}\right|^{-1}$) changes to the power of the
inverse distance.

As an illustration one can consider the simple parabolic point $\mathbf{k}_{0}=\mathbf{k}_{n}^{0}$
in the vicinity of which the function $h\left(\xi_{1},\xi_{2}\right)$
has the expansion \citep{Kossevich1999}: \begin{equation}
h\left(\xi_{1},\xi_{2}\right)=\mathbf{k}_{0}\cdot\widehat{\mathbf{x}}_{0}+\frac{1}{2}\alpha\xi_{1}^{2}+\frac{1}{6}\beta\xi_{2}^{3},\:\alpha=\alpha_{11}^{0},\:\beta=\beta_{111}^{0},\label{math:self:hParabolic}\end{equation}
 where $\widehat{\mathbf{x}}_{0}$ is the unit vector in the direction
normal to the iso-frequency surface at the parabolic point $\mathbf{k}_{0}$.
The local curvilinear coordinates $\xi_{i}$ has the origin at the
parabolic point $\mathbf{k}_{0}$, with the coordinates $\xi_{1}$
and $\xi_{2}$ aligned along the directions of the principal curvatures
of the iso-frequency surface at this point and with the coordinate
$\xi_{3}$ aligned along $\widehat{\mathbf{x}}_{0}$. For the parabolic
point $\mathbf{k}_{0}$ (\ref{math:self:hParabolic}) one of the principal
curvature vanishes ($\alpha_{22}^{0}=0$), while another principal
curvature remain non-zero. Using the expansion (\ref{math:self:hParabolic})
the asymptotic form of the intensity (\ref{math:self:FnkAsymptotic2D})
is given by:\begin{widetext}\begin{equation}
I(\mathbf{r},\omega)\approx\left|\frac{\omega_{0}}{8\pi^{2}\varepsilon_{0}}\frac{\left(\mathbf{a}_{0}^{*}(\mathbf{r}_{0})\cdot\mathbf{d}\right)\mathbf{a}_{0}(\mathbf{r})}{\left|\mathbf{V}_{0}\right|}e^{i\mathbf{k}_{0}\mathbf{\cdot x}}\oint_{-\infty}^{\infty}d\xi_{1}d\xi_{2}\exp\left(i\left|\mathbf{x}\right|\left(\frac{\alpha}{2}\xi_{1}^{2}+\frac{\beta}{6}\xi_{2}^{3}\right)\right)\right|^{2}\label{math:self:FnParabolic}\end{equation}
where $\mathbf{A}_{0}(\mathbf{r})=\mathbf{a}_{0}(\mathbf{r})e^{i\mathbf{k}_{0}\mathbf{\cdot r}}$
and $\mathbf{V}_{0}$ are the Bloch mode and the group velocity associated
with the parabolic point $\mathbf{k}_{0}$, respectively. Calculating
integrals in (\ref{math:self:FnParabolic}), leads to\begin{equation}
\int_{-\infty}^{\infty}d\xi\exp\left(i\frac{x\alpha}{2}\xi^{2}\right)=\sqrt{\frac{2\pi}{x\left|\alpha\right|}}\exp\left(-\frac{i\pi}{4}\mathrm{sing}\left(\alpha\right)\right)\label{math:self:integral1}\end{equation}
 for the direction $\xi_{1}$ and \begin{equation}
\int_{-\infty}^{\infty}d\xi\exp\left(i\frac{x\beta}{6}\xi^{3}\right)=\frac{3}{^{3}\sqrt{x\left|\beta\right|}}\Gamma\left(\frac{4}{3}\right),\label{math:self:integral2}\end{equation}
 for direction $\xi_{2}$. Here $\Gamma\left(\frac{4}{3}\right)$
is the Gamma function. Now, combining (\ref{math:self:integral1})
and (\ref{math:self:integral2}) the following expression for the
asymptotic vector potential associated with the parabolic point (\ref{math:self:hParabolic})
can be obtained \citep{maradudin64,maris1983}:\begin{equation}
I(\mathbf{r},\omega)\approx\left|\frac{3}{2^{5/2}}\frac{\omega_{0}}{\pi^{3/2}\varepsilon_{0}}\exp\left(\frac{\pi}{4}\left(\mathrm{sign}(\alpha)\right)\right)\Gamma\left(\frac{4}{3}\right)\frac{\left(\mathbf{A}_{0}^{*}(\mathbf{r}_{0})\cdot\mathbf{d}\right)\mathbf{A}_{0}(\mathbf{r})}{\left|\mathbf{V}_{0}\right|}\frac{1}{\left|\alpha\right|^{1/2}\left|\beta\right|^{1/3}\left|\mathbf{r}-\mathbf{r}_{0}\right|^{5/6}}\right|^{2}.\label{math:self:FnAsymParabolic}\end{equation}
\end{widetext}

The emission intensity associated with a parabolic point falls off
with the distance as $\left|\mathbf{x}\right|^{-5/3}$ in contrast
to the usual inverse square law $\left|\mathbf{x}\right|^{-2}$ for
other directions. If there are no additional singularities on the
parabolic line, the product $\left|\alpha\right|\left|\beta\right|^{1/2}\sim K$,
where $K$ is the Gaussian curvature at an arbitrary point of the
iso-frequency surface. Then the emission intensity is proportional
to $K^{-1}\left|\beta\right|^{-1/6}$. Thus, at large $\left|\mathbf{x}\right|$,
the energy flux along the direction corresponding to a parabolic point
on the iso-frequency surface exceeds the energy flux along the direction
corresponding to an elliptical point in the ratio $\left|\mathbf{x}\right|^{1/3}\left|\beta\right|^{1/6}$.

The expression (\ref{math:self:FnAsymParabolic}) gives the asymptotic
emitted intensity in the direction $\widehat{\mathbf{x}}_{0}$ associated
with a parabolic point on the iso-frequency surface $\mathbf{k}_{0}$,
so in the direction of the group velocity at the parabolic point.
Now, the asymptotic intensity for directions $\widehat{\mathbf{x}}$
near the direction of that group velocity will be calculated. As before,
the origin of the coordinates $\xi_{i}$ is chosen at the parabolic
point, where the direction of observation $\widehat{\mathbf{x}}$
is coincides with direction $\widehat{\mathbf{x}}_{0}$. It is assumed,
that the principal curvature vanishes in the $\xi_{2}$ direction.
Let the position $\mathbf{x}$ ($\mathbf{x}\Vert\widehat{\mathbf{x}}$)
be described by coordinates $x_{i}$. Then, since $\mathbf{x}$ is
nearly parallel to $\widehat{\mathbf{x}}_{0}$ one have from (\ref{math:self:hParabolic})
\citep{Kossevich1999}:\[
\xi_{3}=\frac{1}{2}\alpha\xi_{1}^{2}+\frac{1}{6}\beta\xi_{2}^{3}\]
 and \begin{equation}
\mathbf{k}_{n}\cdot\mathbf{x}\approx\mathbf{k}_{0}\cdot\mathbf{x}+\xi_{1}x_{1}+\xi_{2}x_{2}+\left(\frac{1}{2}\alpha\xi_{1}^{2}+\frac{1}{6}\beta\xi_{2}^{3}\right)x_{3}.\label{math:self:kxParabolic}\end{equation}
 Using expansion (\ref{math:self:kxParabolic}) the asymptotic form
of the intensity (\ref{math:self:FnkAsymptotic2D}) is given by\begin{widetext}\begin{equation}
I(\mathbf{r},\omega)\approx\left|\frac{\omega_{0}}{8\pi^{2}\varepsilon_{0}}\frac{\left(\mathbf{a}_{0}^{*}(\mathbf{r}_{0})\cdot\mathbf{d}\right)\mathbf{a}_{0}(\mathbf{r})}{\left|\mathbf{V}_{0}\right|}e^{i\mathbf{k}_{0}\cdot\mathbf{x}}\int_{-\infty}^{\infty}d\xi_{1}e^{i\left(x_{1}\xi_{1}+\frac{1}{2}\alpha\left|\mathbf{x}\right|\xi_{1}^{2}\right)}\int_{-\infty}^{\infty}d\xi_{2}e^{i\left(x_{2}\xi_{2}+\frac{1}{6}\beta\left|\mathbf{x}\right|\xi_{2}^{3}\right)}\right|^{2}\label{math:self:FnParabolicAiry}\end{equation}
where the fact that $\left|\mathbf{x}\right|$ and $x_{3}$ are approximately
equal was used. The integral in (\ref{math:self:FnParabolicAiry})
over $\xi_{1}$ is calculated simply to be\[
\int_{-\infty}^{\infty}d\xi_{1}e^{i\left(x_{1}\xi_{1}+\frac{1}{2}\alpha\left|\mathbf{x}\right|\xi_{1}^{2}\right)}=\frac{\sqrt{2\pi}}{\left|\alpha\right|^{1/2}\left|\mathbf{x}\right|^{1/2}}e^{-i\frac{x_{1}^{2}}{2\alpha\left|\mathbf{x}\right|}}e^{-\frac{i\pi}{4}\mathrm{sign}\left(\alpha\right)},\]
 while the integral over $\xi_{2}$ results in\[
\int_{-\infty}^{\infty}d\xi_{2}e^{i\left(x_{2}\xi_{2}+\frac{1}{6}\beta\left|\mathbf{x}\right|\xi_{2}^{3}\right)}=\frac{2^{4/3}\pi}{\left|\beta\right|^{1/3}\left|\mathbf{x}\right|^{1/3}}\mathrm{Ai}\left(x_{2}\frac{2^{1/3}}{\left|\beta\right|^{1/3}\left|\mathbf{x}\right|^{1/3}}\right),\]
 where Ai is the Airy function. Then the asymptotic emitted intensity
(\ref{math:self:FnParabolicAiry}) is finally given by\begin{equation}
I(\mathbf{r},\omega)\approx\left|\frac{1}{2^{7/6}}\frac{\omega_{0}}{\pi\varepsilon_{0}}\exp\left(-i\left(\frac{x_{1}^{2}}{2\alpha\left|\mathbf{x}\right|}+\frac{\pi}{4}\mathrm{sign}\left(\alpha\right)\right)\right)\frac{\left(\mathbf{A}_{0}^{*}(\mathbf{r}_{0})\cdot\mathbf{d}\right)\mathbf{A}_{0}(\mathbf{r})}{\left|\mathbf{V}_{0}\right|}\frac{1}{\left|\alpha\right|^{1/2}\left|\beta\right|^{1/3}\left|\mathbf{r}-\mathbf{r}_{0}\right|^{5/6}}\mathrm{Ai}\left(\frac{x_{2}}{a}\right)\right|^{2},\label{math:self:FnAsymAiry}\end{equation}
 where $a=\left(\left|\beta\right|\left|\mathbf{x}\right|/2\right)^{1/3}$.
\end{widetext}

\begin{figure}[tb]

\begin{centering}
\includegraphics[width=0.9\columnwidth]{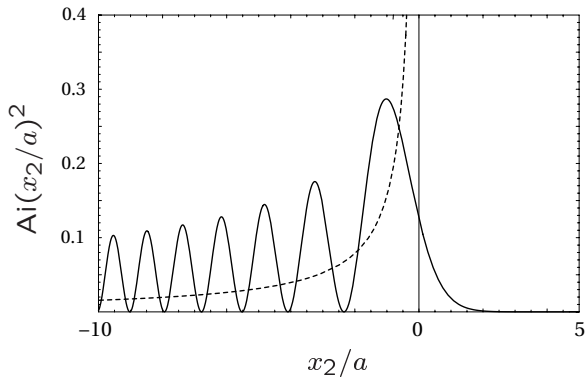}
\par\end{centering}

\caption{\label{fig:self:Airy}Plot of Airy function $\left[\mathrm{Ai}\left(x_{2}/a\right)\right]^{2}$
(solid line) and its mean value $\left(1/4\pi\right)^{-1}\left(x_{2}/a\right)^{-1/2}$
(dashed line). The dashed line can be considered as the {}``geometrical
optics'' approximation of the Airy function.}

\end{figure}

As in the case of the asymptotic intensity (\ref{math:self:FnAsymParabolic}),
energy flux in the direction $\mathbf{x}$ falls off as $\left|\mathbf{x}\right|^{-5/3}$
and exceeds the energy flux along the other directions in the ratio
$\left|\mathbf{x}\right|^{1/3}\left|\beta\right|^{1/6}$. The dependence
of the energy flux in the plane of the vanishing principal curvature
is given by the square of the Airy function $\left[\mathrm{Ai}\left(x_{2}/a\right)\right]^{2}$.
When $x_{2}/a$ is positive, the energy flux is small and exponentially
drops while the angle between direction of observation and direction
corresponded to the parabolic point increases (Fig. \ref{fig:self:Airy}).
For negative $x_{2}/a$, the flux oscillates rapidly and has a mean
value averaged over one cycle proportional to $\sim\left(x_{2}/a\right)^{-1/2}$
(Fig.\ref{fig:self:Airy}). A mean value of the energy flux is then
proportional to $\left|\alpha\right|^{-1}\left|\beta\right|^{-1/2}\sim K^{-1}$
and $\left|\mathbf{x}\right|^{-2}$ and coincides with the asymptotic
energy flux associated with an elliptical point of iso-frequency surface
(\ref{math:self:FnAsymGeometrical}), demonstrating focusing of the
energy flux in the direction corresponding to the parabolic point
of the iso-frequency surface.

\section{\label{sec:self:physics}Interference of Bloch eigenwaves}

\begin{figure}[tb]

\begin{centering}
\includegraphics[width=0.9\columnwidth]{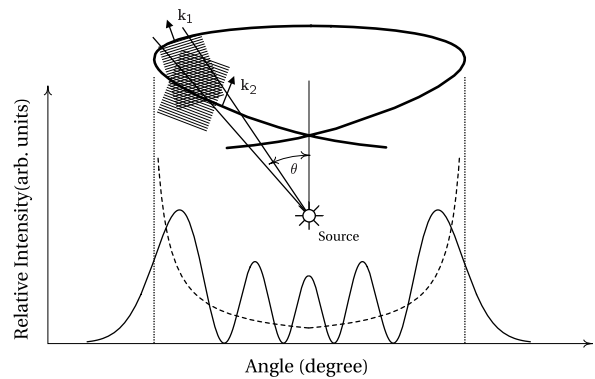}
\par\end{centering}

\caption{\label{fig:self:oscillation}Diagram illustrating the self-interference
of the Bloch eigenwaves in a photonic crystal. A section of the wave
surface with a fold is presented (thick solid line). The asymptotic
intensity (solid line) displays oscillations as the angle passes the
fold section of the wave surface. The asymptotic intensity in the
{}``geometrical optics'' limit demonstrates focusing caustics in
the direction of the folds (dashed line). Within a small solid angle
the far-field of a point source consists on superposition of three
Bloch eigenwaves with different wave vectors. These Bloch eigenwaves
interfere leading to oscillations in the intensity distribution. Only
two wave vectors are illustrated for clarity.}

\end{figure}

The vanishing curvature of the iso-frequency surface results in the
folds of the wave front (wave surface) \citep{chigrin2004}. Then,
for the direction near the fold of the wave surface the field is a
superposition of several Bloch eigenwaves (Fig.~\ref{fig:self:oscillation}).
In the far-field, where the source-to-detector distance is much larger
than the source size and the wavelength, the part of the wave front
limited to the small solid angle can be approximated as a Bloch eigenwave
with the group velocity within this angle. If there is a relative
difference in the lengths or directions of the wave vectors of Bloch
eigenwaves, the eigenwaves can interfere yielding an oscillations
in the energy flux distribution. This can be already seen from the
general expression for the emitted intensity (\ref{eq:I(r,w)}).

Two general conditions are required for the interference to occur.
The polarization states of the Bloch eigenwaves must be nonorthogonal
and the Bloch eigenwaves must overlap in space \citep{russell86}.
This kind of interference of the Bloch eigenwaves will be called further
\emph{a self-interference}, to stress that the field produced by the
light source inside a photonic crystal can interfere with itself producing
an interference patten in the energy flux distribution. A similar
self-interference effect also happens in the case of ballistic phonons
propagation in an acoustically anisotropic crystals \citep{maris1983,wolfe92}.

\begin{figure}[tb]

\begin{centering}
\includegraphics[width=0.9\columnwidth]{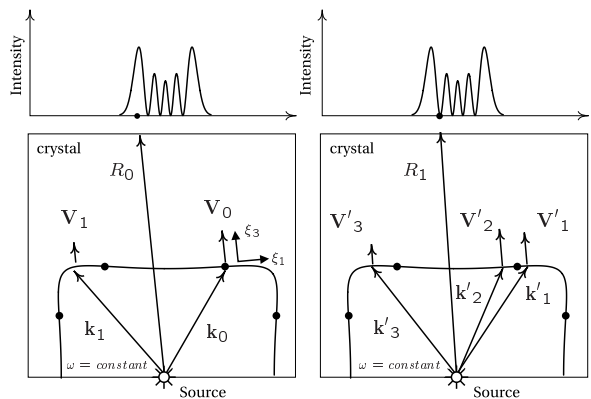}
\par\end{centering}

\caption{\label{fig:self:interference}Schematic of self-interference in a
photonic crystal. The iso-frequency surface is a sketch of the iso-frequency
surface of a real 2D photonic crystal. The plot on the top is the
detected intensity distribution. Left: For the wave vector $\mathbf{k}_{0}$,
the group velocity is $\mathbf{V}_{0}$ and the wave arrives at $\mathbf{R}_{0}$.
Right: There are two eigenwaves with different wave vectors $\mathbf{k}_{i}'$,
which group velocity points in the same direction $\mathbf{R}_{1}$.
The difference in the wave vectors of these eigenwaves results in
the waves arriving at $\mathbf{R}_{1}$ with different phases and
leads to interference.}

\end{figure}

For a more qualitative measure of the self-interference effect the
iso-frequency surface superimposed on a photonic crystal is considered
in figure \ref{fig:self:interference}. Here evaluation presented
by Hauser et al. \citep{wolfe92} for the self-interference of ultrasound
in a crystal is followed. In the figure~\ref{fig:self:interference},
dots are parabolic points of zero curvature. The light source is located
near the bottom surface of the photonic crystal and generates uniform
distribution of wave vectors. The Bloch eigenwave with wave vector
$\mathbf{k}_{0}$ propagates with the group velocity $\mathbf{V}_{0}$,
normal to the iso-frequency surface at $\mathbf{k}_{0}$, arriving
at the point $\mathbf{R}_{0}$ on the opposite surface of the crystal.
Near the parabolic point the iso-frequency surface are practically
flat, neighboring wave vectors have nearly the same group velocity.
This gives rise to the high-intensity caustic in the detected intensity
distribution (Fig.~\ref{fig:self:interference}-left). If the detector
is moved to a point $\mathbf{R}_{1}$ slightly away from $\mathbf{R}_{0}$,
two distinct Bloch eigenwaves with different wave vectors $\mathbf{k}'_{1}$
and $\mathbf{k}'_{2}$ near $\mathbf{k}_{0}$ arrive at the detector
(Fig.~\ref{fig:self:interference}-right). If the surface were perfectly
flat near $\mathbf{k}_{0}$, then $\mathbf{k}'_{1}\cdot\mathbf{R}_{1}=\mathbf{k}'_{2}\cdot\mathbf{R}_{1}$,
and the two eigenwaves would always remain in phase at the detector,
interfering constructively. In reality the iso-frequency surface is
curved near the parabolic point, so as $\mathbf{R}_{1}$ is rotated
downward the corresponding waves begin to interfere destructively,
producing an Airy pattern (Figs.~\ref{fig:self:Airy}-\ref{fig:self:interference}).
If $\mathbf{k}'_{1}$ and $\mathbf{k}'_{2}$ are close two $\mathbf{k}_{0}$,
and that $\mathbf{q}\equiv\mathbf{k}'_{1}-\mathbf{k}_{0}\approx\mathbf{k}'_{2}-\mathbf{k}_{0}$,
then destructive interference will take place, if the total phase
difference of the light as it travels through the sample, $2\mathbf{q}\cdot\mathbf{R}_{1}$,
is an odd integer multiple of $\pi$.

Strictly speaking, there is one more Bloch eigenwave following in
the observation direction in the fold region of the wave surface (Figs.~\ref{fig:self:oscillation}-\ref{fig:self:interference}).
This eigenwave is depicted as $\mathbf{V}_{1}$ and $\mathbf{V}'_{3}$
in the left and right panels of figure~\ref{fig:self:interference},
respectively. To obtain a complete picture of the self-interference
near the fold of the wave surface, a three-wave interference should
be taken into account, which would lead to more complicated interference
patterns in the intensity distribution. Here, the influence of this
third eigenwave is neglected for simplicity, hence this eigenwave
usually has a relatively small group velocity and could arrive at
the detector too late to interfere with the eigenwaves $\mathbf{k}'_{1}$
and $\mathbf{k}'_{2}$.

From the perspective of the self-interference effect, the mean value
of the asymptotic energy flux (\ref{math:self:FnAsymAiry}) averaged
over one cycle of Airy oscillations can be viewed as a {}``geometrical
optics'' approximation of the actual energy flux. This approximation
then corresponds to the ray description of wave propagation, where
the energy flux is simply proportional to the density of rays crossing
a detector surface. In this picture the interference among different
rays is neglected. Then, the emission rate enhancement in the focusing
direction can also be interpreted as the relative increase of the
rays density or as an increased probability of the photon emission
in this observation direction. As it has been mentioned above, the
mean value of the asymptotic flux (\ref{math:self:FnAsymAiry}) coincides
with the asymptotic energy flux (\ref{math:self:FnAsymGeometrical})
derived for elliptical points of the iso-frequency surface. So, the
asymptotic energy flux (\ref{math:self:FnAsymGeometrical}) corresponded
to an elliptical point can be also considered as a {}``geometrical
optics'' approximation, and can be used for all points of the iso-frequency
surface within this approximation.

In a typical experiment the differences between the energy flux (\ref{math:self:FnAsymAiry})
and its {}``geometrical optics'' approximation (\ref{math:self:FnAsymGeometrical})
will be reduced by the effect of the finite size of the light source
and the detector. It is clear that if the linear dimensions of the
source area and detector are $L$, the intensity is averaged over
$x_{2}$ values with a spread of $L$. To see the oscillations of
the energy flux one therefore needs $L\leq\Delta\theta\times R_{1}$,
where $R_{1}$ is the distance between the source and the detector
and $\Delta\theta$ is an angular separation of the fringes of intensity
distribution.

To estimate this angular separation, Bloch eigenwaves $\mathbf{k}'_{1}$
and $\mathbf{k}'_{2}$ are further approximated by plane waves. Then
their superposition at the detector position $\mathbf{R}_{1}$, assuming
that they have the same polarization, is \citep{wolfe92}:\[
e^{i\mathbf{k}'_{1}\cdot\mathbf{R}_{1}}+e^{i\mathbf{k}'_{2}\cdot\mathbf{R}_{1}}=2\cos\left(\Delta\mathbf{k}\cdot\mathbf{R}_{1}\right)e^{i\mathbf{k}_{0}\cdot\mathbf{R}_{1}},\]
 which is a plane wave with average wave vector $\mathbf{k}_{0}=\left(\mathbf{k}'_{1}+\mathbf{k}'_{2}\right)/2$,
modulated by a cosine function with effective wave vector $\Delta\mathbf{k}/2=\left(\mathbf{k}'_{2}-\mathbf{k}'_{1}\right)/2$.
When $\Delta\mathbf{k}\cdot\mathbf{R}_{1}=\Delta k_{\Vert}R_{1}=\pi,$
the waves interfere destructively at the detector. To estimate $\Delta k_{\Vert}$
a local Cartesian coordinate system $\xi_{i}$ with the origin at
$\mathbf{k}_{0}$ and $\xi_{3}$ along $\mathbf{V}_{0}$ is chosen
as it is shown in the figure~\ref{fig:self:interference}-left. Then,
the iso-frequency surface near the parabolic point can by parametrized
as $\xi_{3}=-a\xi_{2}^{3}/k_{0}$ and $\Delta k_{\Vert}=-2\xi_{3}=2a\xi_{2}^{3}/k_{0}^{2}$
\citep{wolfe92}. Therefore, the first minimum in the intensity will
occur when\[
\xi_{2}=\left(\pi k_{0}^{2}/2aR_{1}\right)^{1/3}\sim R_{1}^{-1/3}\lambda^{-2/3},\]
 where $\lambda=2\pi/k_{0}$ is the average wavelength. Finally, the
coordinate-space angle between the intensity maximum and the first
minimum is given by \citep{wolfe92}:\begin{equation}
\Delta\theta\approx\left|\mathbf{V}_{1}'-\mathbf{V}_{0}\right|/\mathbf{V}_{0}=3a\left(\pi/2ak_{0}R_{1}\right)^{2/3}\sim\left(\lambda/R_{1}\right)^{2/3}.\label{math:self:angleAiry}\end{equation}
 Then for optical wavelength, e.g., 500 nm, and a distance to the
detector of 1 cm the linear dimension of the light source and the
spatial resolution of the detector should be smaller than 10 $\mu$m.
So, in most experiments the {}``geometrical optics'' approximation
(\ref{math:self:FnAsymGeometrical}) reasonably represents an asymptotic
emission intensity of the light source inside a photonic crystal.

\begin{figure}[!tb]

\begin{centering}
\includegraphics[width=0.9\columnwidth,keepaspectratio]{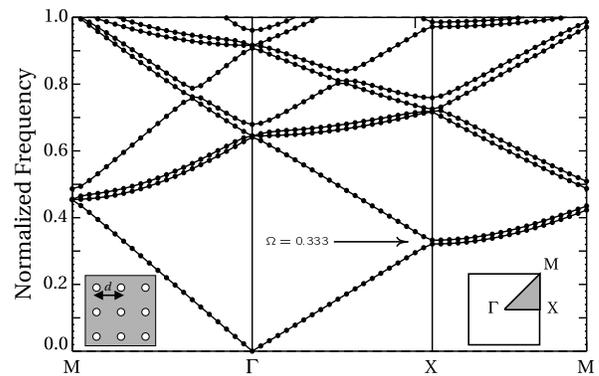}
\par\end{centering}

\caption{Photonic band structure of the square lattice of air holes made in
a polymer. Polymer has the refractive index 1.56. Radius of holes
is $r=0.15d$, where $d$ is the lattice period. The band structure
is given for TM polarization. The frequency is normalized to $\Omega=\omega d/2\pi c=d/\lambda$.
$c$ is the speed of light in the vacuum. Insets show the first Brillouin
zone (right) and a part of the lattice (left).\label{fig:self:pbs}}

\end{figure}

\section{\label{sec:self:polymer}Numerical example}

In this section different approximations of the light emission pattern,
discussed in section \ref{sec:self:asymptotic}, are compared. Numerical
calculations are done for a point source placed inside a two-dimensional
polymer photonic crystal. A point source produces an isotropic and
uniform distribution of wave vectors $\mathbf{k}_{n}$ with the frequency
$\omega_{0}$. Then, the asymptotic field in (\ref{math:self:FnkAsymptotic2D})
and (\ref{math:self:FnAsymGeometrical}) should be averaged over the
dipole moment orientation, which yields a factor of $\left|\mathbf{d}\right|/3$.

As it was pointed out in section \ref{sec:self:asymptotic}, the main
contribution to the far-field of a point source inside a photonic
crystal comes from the vicinity of the wave vector of the eigenmodes
with the group velocity in the observation direction. That means that
the far-field emission intensity of a point source is mainly given
by the square of the integral in (\ref{math:self:FnkAsymptotic2D})
\begin{equation}
I_{w}\sim\left|\oint_{-\infty}^{\infty}d^{2}\mathbf{k}_{n}e^{i\mathbf{k}_{n}(\mathbf{r}-\mathbf{r}_{0})}\right|^{2}.\label{math:self:Integral}\end{equation}
In what follows, the contribution only from one photonic band is considered.
In the {}``geometrical optics'' approximation (\ref{math:self:FnAsymGeometrical})
the main contribution to the far-field emission intensity is then
given by an inverse Gaussian curvature of the iso-frequency surface,
\begin{equation}
I_{g}\sim\sum_{\nu}\left|K_{n\mathbf{k}}^{\nu}\right|.^{-1}\label{math:self:Curvature}\end{equation}

\begin{figure}[!tb]

\begin{centering}
\includegraphics[width=0.7\columnwidth,keepaspectratio]{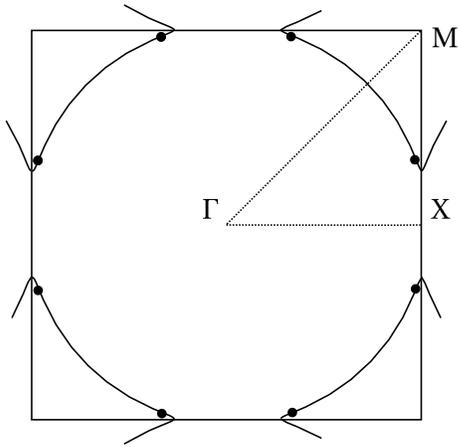}
\par\end{centering}

\caption{Iso-frequency contour of the square lattice photonic crystal (Fig.
\ref{fig:self:pbs}) for the normalized frequency $\Omega=0.333$.
Parabolic point are marked by the black dots. The first Brillouin
zone of the lattice is plotted in order to show the spatial relation
between zone boundary and iso-frequency contours.\label{fig:self:IFS-0.333}}

\end{figure}

\begin{figure}[!tb]

\begin{centering}
\includegraphics[width=0.8\columnwidth]{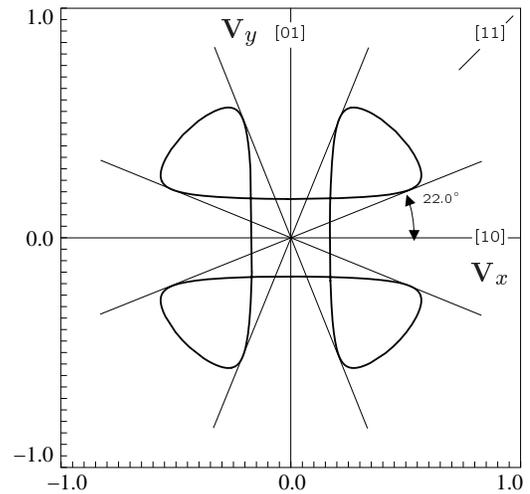}
\par\end{centering}

\caption{Wave contour corresponded to the normalized frequency $\Omega=0.333$.
The group velocity is plotted in the units of the speed of light in
vacuum. The directions corresponded to the folds of the wave contour
are shown.\label{fig:self:WC-0.333}}

\end{figure}

\begin{figure}[!tb]

\begin{centering}
\includegraphics[width=0.8\columnwidth]{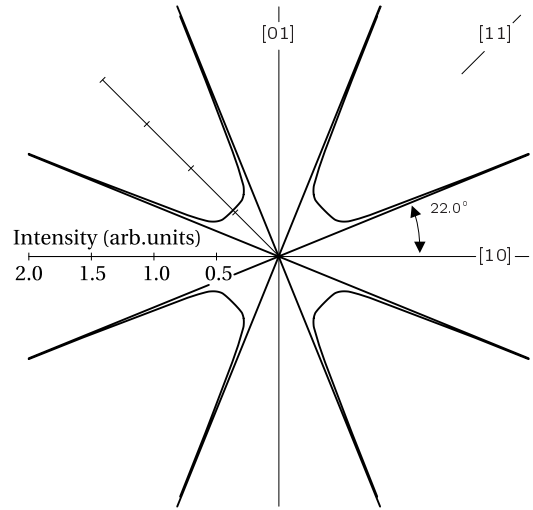}
\par\end{centering}

\caption{Angular distribution of radiative power corresponding to the normalized
frequency $\Omega=0.333$. The directions of infinite radiative power
(caustic) coincide with the directions of the folds of the wave contour
(Fig.~\ref{fig:self:WC-0.333}).\label{fig:self:RP-0.333}}

\end{figure}

\begin{figure}[!tb]

\begin{centering}
\includegraphics[width=0.8\columnwidth,keepaspectratio]{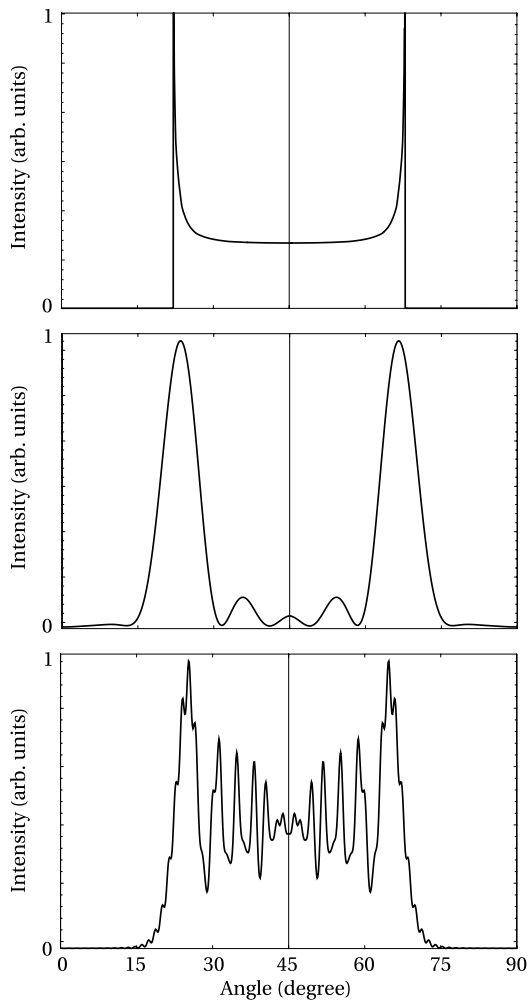}
\par\end{centering}

\caption{{}``Geometrical optics'' (\ref{math:self:Curvature}) and {}``wave
optics'' (\ref{math:self:Integral}) approximations of the far-field
emission intensity. Normalized frequency is $\Omega=0.333$. The distance
between a point source and a detector is 100 lattice periods. The
top panel is for the {}``geometrical optics'' approximation. The
middle and bottom panels are for the {}``wave optics'' approximations.
See text for more details. \label{fig:self:compare}}

\end{figure}

The angular distribution of the far-field intensity depends on the
topology of the iso-frequency surface of the crystal at the emission
frequency (\ref{math:self:Integral}-\ref{math:self:Curvature}).
In what follows, an infinite two-dimensional square lattice of air
holes in a polymer background is considered. Polymer has the refractive
index $n=1.56$, radius of holes is $r=0.15d$, where $d$ is the
lattice period. The consideration is limited to the in-plane propagation
of the TM mode of the crystal. The photonic band structure of such
a photonic crystal is presented in the figure \ref{fig:self:pbs}.
The band structure has been calculated using the plane wave expansion
method \citep{johnson2001}.

The iso-frequency contours for the normalized frequency $\Omega=0.333$
is presented in figure \ref{fig:self:IFS-0.333}. The frequency belongs
to the first photonic band and it is within the first stopband in
the $\Gamma\mathrm{X}$ direction of the crystal. To plot an iso-frequency
contour, the photonic band structure for all wave vectors within the
irreducible Brillouin zone was calculated and then an equation $\omega(\mathbf{k})=\omega_{0}$
was solved for a given frequency $\omega_{0}$. The iso-frequency
contour is an \emph{open} contour and has alternating regions of the
Gaussian curvature with a different sign. Parabolic points, where
the Gaussian curvature vanishes, are marked by black dots in the figure
\ref{fig:self:IFS-0.333}. The vanishing curvature results in the
folds of the wave contour and in the focusing of the light in the
folds direction \citep{chigrin2004}. The wave contour corresponding
to the iso-frequency $\Omega=0.333$ is presented in the figure \ref{fig:self:WC-0.333}.
A pair of the parabolic points in the first quarter of the Brillouin
zone results in a cuspidal structure of the wave contours in the first
quarter of the coordinate space. In the figure \ref{fig:self:RP-0.333}
the polar plot of the main contribution to the far-field intensity
in the {}``geometrical optics'' approximation (\ref{math:self:Curvature})
is presented. The energy flux is strongly anisotropic, showing relatively
small intensity in the directions of the stopband, and infinite intensity
(caustics) in the directions of the folds.

\begin{figure}[!tb]

\begin{centering}
\includegraphics[width=0.8\columnwidth,keepaspectratio]{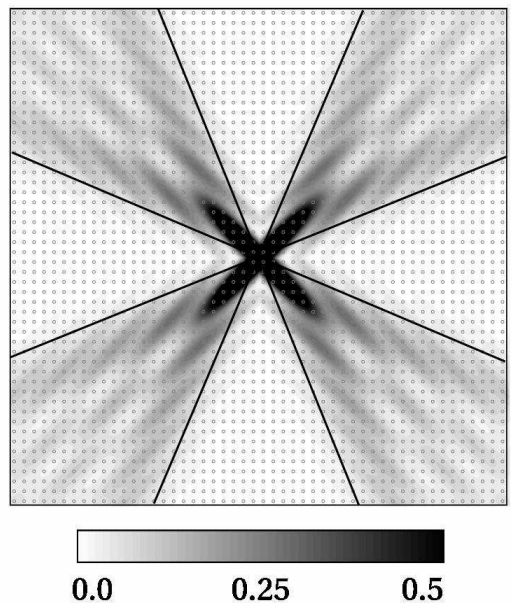}
\par\end{centering}

\caption{An asymptotic map of the intensity distribution inside a $50\times50$
photonic crystal. Normalized frequency is $\Omega=0.333$. The structure
of the crystal is superimposed on the field map. Folds directions
are shown by black lines.\label{fig:self:2Dmap}}

\end{figure}

\begin{figure}[!tb]

\begin{centering}
\includegraphics[width=0.8\columnwidth,keepaspectratio]{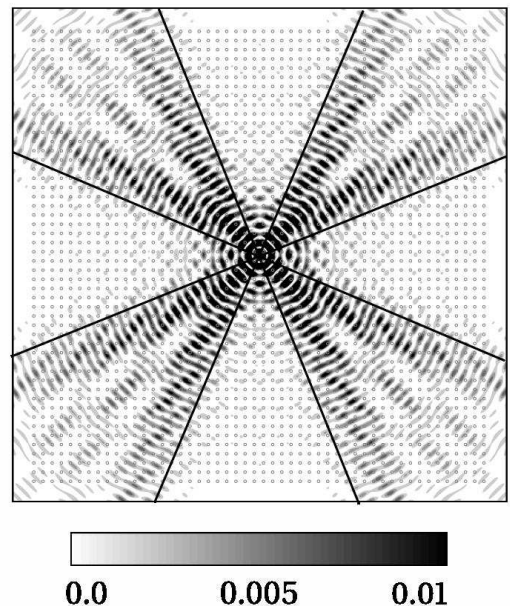}
\par\end{centering}

\caption{FDTD calculation. Map of the modulus of the Poynting vector field
for a $50\times50$ rod photonic crystal excited by a point isotropic
source with the normalized frequency $\Omega=0.333$. The asymptotic
directions of photon focusing caustics are shown as black lines.\label{fig:self:fdtd}}

\end{figure}

\begin{figure}[!tb]

\begin{centering}
\includegraphics[width=0.95\columnwidth,keepaspectratio]{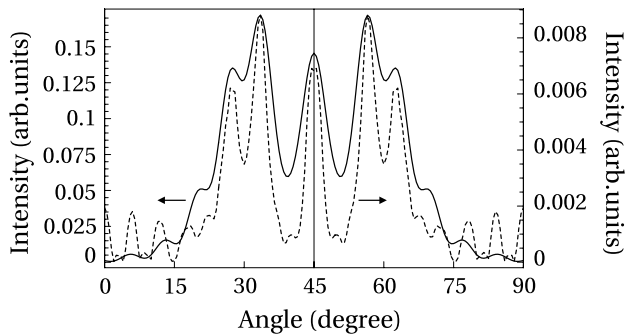}
\par\end{centering}

\caption{The comparison of the intensity distribution for the asymptotic (solid
line) and the FDTD (dashed line) calculations. The distance between
the point source and the detector is 20 lattice periods. \label{fig:self:FDTD-Asymptotic}}

\end{figure}

In figure \ref{fig:self:compare} the comparison of the {}``geometrical
optics'' (\ref{math:self:Curvature}) and {}``wave optics'' (\ref{math:self:Integral})
approximations of the far-field emission intensity is given. A normalized
inverse Gaussian curvature of the iso-frequency surface $\Omega=0.333$
is presented in the top panel. Focusing directions, $22^{\circ}$
apart from the {[}10] direction of the square lattice, are clearly
seen. The integral (\ref{math:self:Integral}) evaluated for the distance
100 period apart from the point source is given in the middle and
the bottom panels of figure \ref{fig:self:compare}. The normalized
intensity distribution presented in the middle panel was calculated
by reducing the integration limits in (\ref{math:self:Integral})
to the close neighborhood of the parabolic point of the iso-frequency
surface. Then the result of the integration is similar to one in (\ref{math:self:FnAsymAiry})
and an angular distribution of the emission intensity resembles the
square of the Airy function. Actually, this approximation takes into
account an interference of only two Bloch eigenwaves in the fold region.
If the three wave interference is taken into account, by extending
the integration limits in (\ref{math:self:Integral}) over all iso-frequency
surface in the first quarter of the Brillouin zone, a more complex
interference pattern appears in the angular emission intensity distribution
(Fig. \ref{fig:self:compare}-bottom). Both {}``wave optics'' approximations
show an intensity enhancement along fold directions.

In figure \ref{fig:self:2Dmap} a two dimensional map of the intensity
distribution inside a $50\times50$ photonic crystal is presented.
The intensity distribution was calculated using Eq. (\ref{math:self:Integral})
by integration over complete iso-frequency contour $\Omega=0.333$.
The structure of the crystal is superimposed on the field map. Folds
directions are shown by black lines. The focusing of the light in
the fold direction together with an Airy-like oscillations between
folds directions are clearly seen in figure \ref{fig:self:2Dmap}.

To substantiate an asymptotic analysis, the finite difference time
domain (FDTD) calculations was done \citep{taflove95,sullivan02}.
The simulated structure was a $50\times50$ lattice. The crystal is
surrounded by an extra $2d$ wide layer of polymer. The simulation
domain was discretized into squares with a side $\Delta=d/32$. The
total simulation region was $1728\times1728$ cells plus 8-cell wide
perfectly matched layer (PML) \citep{berenger94}. The point isotropic
light source was modeled by a current density source \citep{taflove95,sullivan02}
with a homogeneous spacial dependence and sinusoidal temporal dependence
of the signal.

In figure~\ref{fig:self:fdtd} the map of the modulus of the Poynting
vector field is shown, when the crystal is excited by a point isotropic
source. The point source is placed in the middle of the crystal. A
field map is shown for one instant time step. The snap-shots were
captured after 10000 time steps, where the time step was $4.38\times10^{-17}$
s (0.99 of the Courant value). The structure of the crystal is superimposed
on the field map. One can see, that the emitted light is focused in
the directions of the folds (black lines). Moreover, an interference
pattern between the folds directions is in a reasonable agreement
with the interference pattern predicted using the asymptotic analysis
(Fig. \ref{fig:self:2Dmap}). For the FDTD calculations, a periodic
modulation of the intensity in the radial direction will go away if
time averaging is performed. The comparison of the intensity distribution,
20 periods apart from the point source is given in figure \ref{fig:self:FDTD-Asymptotic}.
A reasonable agreement between interference minima and maxima positions
for the asymptotic (solid line) and the FDTD (dashed line) calculations
is shown. The disagreement in the absolute values of the angular intensity
distributions is mainly because of the prefactor of the integral in
(\ref{math:self:FnkAsymptotic2D}) which was neglected in (\ref{math:self:Integral})
for simplicity.

\section{\label{sec:self:summary}Summary}

It was shown, that the intensity modulation of the angular resolved
emission spectra is not only due to the emission rate modification,
but also is the result of the interference of several photonic crystal
eigenmodes with different wave vectors approaching detector at the
same moment of time. Using an asymptotic analysis of classical Green
function, {}``geometrical optics'' and {}``wave optics'' approximations
of the emitted intensity due to a two-level atom were introduced in
the radiation zone. The physical reasons for the interference pattern
formation and the possibilities of experimental observation of them
were discussed. A numerical example was given in the case of polymer
two-dimensional photonic crystal. It was shown that rigorous FDTD
calculations are in a reasonable agreement with the developed approximate
analysis.

\begin{acknowledgments}
This work was partially supported by the EU-IST project APPTech IST-2000-29321,
the BMBF project PCOC 01 BK 253 and the DFG Research Unit 557.
\end{acknowledgments}

\end{document}